\documentstyle[preprint,aps]{revtex}
\begin{document}
\draft
\title{Effective spatial dimension of extremal non-dilatonic black
 $p$-branes and the description of entropy on the world volume}
\author{ Rong-Gen Cai}
\address{CCAST (World Laboratory), P.O. Box 8730, Beijing 100080, 
China \\ 
and Institute of Theoretical Physics, Academia Sinica, P.O. Box 
2735, Beijing 100080, China}
\maketitle

\begin{abstract}
By investigating the critical behavior appearing at the 
extremal limit of the non-dilatonic, black $p$-branes in
 ($d+p$) dimensions, we find that some critical exponents 
 related to the critical point obey the scaling laws. 
 From the scaling laws we obtain that the effective 
 spatial dimension of the non-dilatonic black holes 
 and black strings is one, and is $p$ for the 
 non-dilatonic black $p$-branes. For the dilatonic 
 black holes and  black $p$-branes, the 
 effective dimension will depend on the parameters 
 in theories. Thus, we give an interpretation why 
 the Bekenstein-Hawking entropy may be given a
  simple world volume interpretation only for 
  the non-dilatonic black $p$-branes.


\end{abstract}

Classical general relativity and the quantum field theory in 
curved spacetime
together provide the temperature and entropy of black holes [1-3].
 Although the Bekenstein-Hawking entropy of black holes can indeed
  be derived in the Euclidean path integral method of quantum gravity 
  under the zero-loop approximation [4], a satisfactory statistical 
  interpretation of the entropy
 is still needed. In the last few months, the important progress 
 towards a microspic understanding of the black hole entropy has 
 been made. Strominger and Vafa [5] considered a class of 
 five-dimensional extremal black holes in string theory. They 
 found that the Bekenstein-Hawking entropy of the black holes 
 agrees with that of BPS soliton bound states with same charges. 
Since then, a lot of papers appear to study this agreement in the
  extremal and near-extremal black holes, black strings and black 
  $p$-branes (some extended objects surrounded by event horizons) [6]. 
  Klebanov and Tseytlin [7] found that there are 
  non-dilatonic $p$-branes whose near-extremal entropy may be 
  explained by free massless fields on the world volume. Gubser, 
  Klebanov, and Peet [6] also got the similar result in the black
   3-branes. But the reason seems to be unclear.

In this Letter, by investigating the critical behavior occurring 
at the extremal limit ($r_-=r_+$) of the non-dilatonic, 
black $p$-branes in ($d+p$) dimensions (the dilaton field is a 
constant throughout spactime), we obtain an effective 
spatial dimension of the black $p$-branes. For the non-dilatonic 
black holes and black strings the effective dimension is one, and 
it is just $p$ for the non-dilatonic black $p$-branes. When the 
dilatonic black holes are obtained by the double-dimensional 
reduction of the non-dilatonic black $p$-branes [8], the effective 
dimension is also $p$. For a general coupling constant $a$ governing 
the interaction strength between the dilaton field and the asymmetric 
tensor field, the effective dimension will depend on the parameters 
in theories. Thus, we give an interpretation 
of the result of Klebanov and Tseytlin [7] that the Bekenstein-Hawking
 entropy may be given a simple world volume interpretation only 
 for the non-dilatonic $p$-branes
  (including the non-dilatonic black holes).

We start with the non-dilatonic, ($d+p$)-dimensional action [8],
\begin{equation}
S_{d+p}=\frac{1}{16\pi}\int d^{(d+p)}x\sqrt{-g}
\left[R-\frac{2}{(d-2)!}F^2_{d-2}\right],
\end{equation}
where $R$ is the scalar curvature and $F_{d-2}$ denotes the
 ($d-2$)-form asymmetric tensor field. Performing the 
 double-dimensional reduction by $p$ dimensions, one has the 
 dilatonic $d$-dimensional action:
\begin{equation}
S_d=\frac{1}{16\pi}\int d^d\sqrt{-g}\left[R-2(\nabla \phi)^2-
\frac{2}{(d-2)!}e^{-2a\phi}F^2_{d-2}\right],
\end{equation}
where $\phi$ is the dilaton field, and the constant $a$ is
\begin{equation}
a=\frac{(d-3)\sqrt{2p}}{\sqrt{(d-2)(d+p-2)}}.
\end{equation}
The magnetically charged black holes in the action (2) are [9]
\begin{eqnarray}
ds^2_d&=&\left[1-\left(\frac{r_+}{r}\right)^{d-3}\right]
\left[1-\left(\frac{r_-}{r}\right)^{d-3}\right]^{1-(d-3)b}dt^2
 \nonumber\\
&+&\left[1-\left(\frac{r_+}{r}\right)^{d-3}\right]^{-1}
\left[1-\left(\frac{r_-}{r}\right)^{d-3}\right]^{b-1}dr^2
 \nonumber\\
&+&r^2\left[1-\left(\frac{r_-}{r}\right)^{d-3}\right]^{b}
d\Omega^2_{d-2}, \nonumber \\
e^{a\phi}&=&\left[1-\left(\frac{r_-}{r}\right)^{d-3}
\right]^{-(d-3)b/2}, \nonumber \\
F_{d-2}&=&Q\varepsilon _{d-2},
\end{eqnarray}
where $\varepsilon _{d-2}$ is the volume form on the unit
 ($d-2$)-sphere, the constant $b$ is
\begin{equation}
b=2p/(d-2)(p+1),
\end{equation}
and the charge $Q$ is related to $r_{\pm}$ by
\begin{equation}
Q^2=\frac{(d-3)(d+p-2)}{2(p+1)}(r_+r_-)^{d-3}.
\end{equation}
Thus, one has the non-dilatonic black $p$-brane solutions in 
the action (1) [8]:
\begin{equation}
ds^2_{d+p}=e^{2m\phi}dy^idy^i + e^{2n\phi}ds^2_d,
\end{equation}
where $i=1,2,\cdots,p$, and
\begin{equation}
m=-\frac{\sqrt{2(d-2)}}{\sqrt{p(d+p-2)}}, \ \ n=-\frac{mp}{d-2}.
\end{equation}
From Eqs. (4)-(8), the Hawking temperature and the Bekenstein-Hawking 
entropy per unit volume of $p$-branes for the black $p$-branes (7)
 are easily obtained:
\begin{eqnarray}
&& T=\frac{d-3}{4\pi r_+}\left [1-\left(\frac{r_-}{r_+}\right)^{d-3}
\right]^{1/(p+1)},\\
&& S=\frac{\Omega _{d-2}}{4}r_+^{d-2}\left[1-\left(\frac{r_-}{r_+}
\right)^{d-3}\right]^{p/(p+1)},
\end{eqnarray}
where $\Omega _{d-2}$ is the volume of the unit ($d-2$)-sphere.
 The ADM mass per
 unit volume of $p$-branes is found to be
\begin{equation}
M=\frac{\Omega _{d-2}}{16\pi}\left[(d-2)r_+^{d-3}+\frac{d-2-p(d-4)}
{p+1}r_-^{d-3}\right],
\end{equation}
which satisfies the first law of thermodynamics,
\begin{equation}
dM=TdS+\Phi dQ,
\end{equation}
where $\Phi=\Omega _{d-2}Q/[4\pi (d-3)r_+^{d-3}]$ is the chemical
 potential corresponding to the conservative charge $Q$. According 
 to the formula $C_Q \equiv (\partial M/\partial T)_Q$, the heat 
capacity per unit volume of $p$-branes is
\begin{eqnarray}
C_Q&=&-\frac{\Omega _{d-2}r_+}{4}\frac{\left[1-
\left(\frac{r_-}{r_+}\right)^{d-3}\right]^{p/(p+1)}}
{\left[1-\frac{p+2d-5}{p+1}\left(\frac{r_-}{r_+}\right)^{d-3}
\right]} \nonumber\\
&\times&
\left[(d-2)r_+^{d-3}-
\frac{d-2-p(d-4)}{p+1}r_-^{d-3}\right].
\end{eqnarray}
When the extremal limit ($r_-=r_+$) is approached, the temperature,
 entropy, and the heat capacity approach zero. When 
$1-(p+2d-5)(r_-/r_+)^{d-3}/(p+1)=0$, the heat capacity diverges, which
 corresponds to the critical point of Davies in Kerr-Newman 
 black holes [10]. 
 
In a self-gravitating thermodynamic system, in general, the 
thermodynamic ensembles are not equivalent [11]. Hence, the 
critical bahavior and stability of the syetem are different 
in the different environments (implying the different ensembles).
 In order to discuss the critical behavior of an isolated black 
$p$-brane, it is reasonable to choose the microcanonical 
ensemble [12-14]. 
In this ensemble, the proper Massieu function, which can describe
 conpletely the equilibrium state of a thermodynamic system, is 
 the entropy of the system. For the black $p$-branes (7), 
 rewriting Eq. (12), one has
\begin{equation}
dS=\beta dM -\varphi dQ,
\end{equation}
where $\beta =T^{-1}$ and $\varphi=\beta \Phi$. Applying the 
fluctuation theory of equilibrium thermodynamic under the specified 
environments [11-14] to the black $p$-branes (7), it follows from 
Eq. (14) that the intrinsic variables $x_i=\{M,Q\}$ and the 
conjugate variables $X_i=\{\beta, -\varphi\}$ in the microcanonical 
ensemble. Thus the eigenvalues 
corresponding to the fluctuation modes $\beta$ and $\varphi$ are
\begin{eqnarray}
&&\lambda _m=\left(\frac{\partial M}{\partial \beta}\right)_Q=
-T^2C_Q,\\
&&\lambda _q=-\left(\frac{\partial Q}{\partial \varphi}\right)_M=
-TK_M,
\end{eqnarray}
respectively, where
\begin{eqnarray}
K_M&\equiv & \beta\left(\frac{\partial Q}{\partial \varphi}
\right)_M \nonumber\\
&=&\frac{4\pi (d-3)r_+^{d-3}}{\Omega _{d-2}}\left[1-\frac{d-2
-p(d-4)}{(p+1)(d-2)}
\left(\frac{r_-}{r_+}\right)^{d-3}\right] \nonumber\\
&&\left \{\left[1+\frac{d-2-p(d-4)}{p+1)(d-2)}\left(\frac{r_-}
{r_+}\right)^{d-3} \right] \right. \nonumber\\
&+& \frac{2}{(d-3)}\left(\frac{r_-}{r_+}\right)^{d-3}\left[
\frac{d-3}{p+1}-
\frac{d-2-p(d-4)}{(p+1)(d-2)} \right. \nonumber\\
&\times&\left. \left(1-\frac{p+d-2}{p+1}\left(\frac{r_-}{r_+}
\right)^{d-3}\right)\right] \nonumber\\
&\times& \left. \left[1-\left(\frac{r_-}{r_+}\right)^{d-3}
\right]^{-1}\right\}^{-1}.
\end{eqnarray}
By using the fluctuation theory, we obtain the
 nonvanishing second moments of fluctuations,
\begin{eqnarray}
&&\langle\delta \beta \delta \beta \rangle=
-k_B\frac{\beta ^2}{C_Q},\ \
\langle\delta \varphi \delta \varphi\rangle=
-k_B\frac{\beta }{K_M},\nonumber\\
&&\langle\delta \beta  \delta \Phi\rangle=
 k_B\frac{\beta \Phi}{C_Q}, \ \
\langle\delta \Phi\delta \Phi\rangle=
-k_B\left(\frac{T}{K_M}+\frac{\Phi^2}{C_Q}\right).
\end{eqnarray}
Obviously, the two eigenvalues approach zero and all of these 
second moments diverge when the extremal limit ($r_-= r_+$) is 
approached. As in the ordinary thermodynamics, the divergence of
second moments means that the extremal limit is a critical point 
and a second-order phase transition
takes place from the extremal to nonextremal black $p$-branes. 
As is well known, the extremal black $p$-braes are very different 
from the nonextremal in many aspects, such as the thermodynamic 
description [15] and geometric structures [16]. In particular, 
it has been shown that the extremal black $p$-branes are 
 supersymmetric and the supersymmetry is absent for the nonextremal 
 black $p$-branes [8,9]. So the occurrence of phase transition 
  are consistent with
 the changes of symmetry. The extremal and nonextremal black 
 $p$-branes
  are two different phases. The extremal black $p$-branes 
are in the disordered phase and the nonextremal black $p$-branes 
in the ordered phase. The order parameters
of the phase transition can be defined as the differences of 
the conjugate variables between the two phases [12,13], such
 as $\eta _{\beta}=\beta _+-\beta _-$ and $\eta _{\varphi}=
 \varphi _+-\varphi _-$ can be regarded as the order parameters
  of  black $p$-branes, where the suffixes ``$+$'' and ``$-$''
 mean that the quantity is taken at the $r_+$ and $r_-$, 
respectively. The second-order derivatives of entropy with respect 
to the intrinsic variables are the inverse eigenvalues,
\begin{eqnarray}
\zeta_m&\equiv &\left (\frac{\partial ^2S}{\partial M^2}\right)_Q=
\lambda _{m}^{-1}=-\frac{\beta ^2}{C_Q},\\
\zeta_q&\equiv&\left (\frac{\partial ^2 S}{\partial Q^2}\right)_M=
\lambda _{q}^{-1}=-\frac{\beta }{K_M}.
\end{eqnarray}
 Correspondingly, we can define the critical exponents of these 
 quantities as follows [17],
\begin{eqnarray}
\zeta _m&\sim&\varepsilon _M^{-\alpha}\ \ \ 
({\rm for\ Q\ fixed}),\nonumber\\
         &\sim&\varepsilon _Q^{-\psi}\ \ \ ({\rm for\ M\ fixed}),\\
\zeta_q & \sim & \varepsilon ^{-\gamma}_M \ \ \  
({\rm for\ Q\ fixed }),      \nonumber \\
          & \sim &\varepsilon ^{-\sigma}_Q \ \ \ ({\rm for\ M\ fixed}),\\
\eta _{\varphi}&\sim &\varepsilon ^{\beta}_M \ \ \ 
({\rm for\ Q\ fixed}),\nonumber\\
      &\sim &\varepsilon ^{\delta ^{-1}}_Q \ \ \ ({\rm for\ M\ fixed}),
\end{eqnarray}
where $\varepsilon _M$ and $\varepsilon _Q$ represent the 
infinitesimal deviations of $M$ and $Q$ from their limit values. 
These critical exponents are found to be
\begin{equation}
\alpha=\psi=\gamma=\sigma=\frac{p+2}{p+1}, \ \
 \beta=\delta ^{-1}=-\frac{1}{p+1}.
\end{equation}
 The critical exponents $\beta $ and $\delta ^{-1}$ are negative, 
which shows the fact that the order parameter $\eta _{\varphi}$ 
diverges at the extremal limit. This is because the critical 
temperature is  zero in this phase transition. It is easy to 
check that these critical
  exponents satisfy the scaling laws of the ``first kind,''
\begin{equation}
\alpha +2\beta +\gamma =2,\ \
\beta(\delta-1)=\gamma, \ \
\psi (\beta +\gamma)=\alpha.
\end{equation}
That scaling laws (25) hold for the black $p$-branes is related to 
the fact that the black $p$-brane entropy (10) is a homogeneous 
function, satisfying 
\begin{equation}
S(\lambda M, \lambda Q)=\lambda ^{(d-2)/(d-3) }S(M,Q),
\end{equation}
where $\lambda$ is a positive constant. On the other hand, 
in an ordinary thermodynamic system, an important physical 
quantity related to phase transitions is the two-point 
correlation function, which has  
generally the form for a large distance [17],
\begin{equation}
G(r)\sim \frac{\exp (-r/\xi)}{r^{\bar{d}-2+\eta}},
\end{equation}
where $\eta$ is the Fisher's exponent, $\bar{d}$ is the effective
 spatial dimension of the system under consideration, and $\xi$ is
  the correlation length and diverges at the critical point. 
Similarly, the critical exponents of 
  the correlation 
  length for black $p$-branes can be defined as:
\begin{eqnarray}
\xi &\sim & \varepsilon ^{-\nu}_M \ \ \ ({\rm for\ Q\ fixed}),
\nonumber\\
    &\sim &\varepsilon ^{-\mu}_Q \ \ \ ({\rm for\ M\ fixed}).
\end{eqnarray}
Combining with those in Eq. (25), these critical exponents form 
the scaling laws of the ``second kind''
\begin{equation}
\nu (2-\eta )=\gamma, \ \ \nu \bar{d}=2-\alpha, \ \ 
\mu (\beta +\gamma)=\nu.
\end{equation}
Because of the absence of quantum theory of gravity, we have not 
yet the correlation function of quantum black holes. 
 Here we use 
the correlation function of scalar fields on the background of 
these black $p$-branes to mimic the one of black $p$-branes (from
 the obtained result below, it seems an appropriate approach to 
study the critical behavior of black holes at the present time).
 From the work of Traschen [18] who studied the behavior of 
 scalar fields on the background of Reissner-Nordstr\"{o}m black
  holes, it is found that the inverse surface gravity of the black
 hole  plays the role of the correlation length of scalar 
fields. For the black $p$-branes, this conclusion holds as 
well. With the help of the surface gravity of black 
$p$-branes (7), we obtain
\begin{equation}
\nu=\mu=\frac{1}{p+1}.
\end{equation}
 Substituting (30) into (29), we find
\begin{equation}
\eta=-p,\ \ \bar{d}=p.
\end{equation}
When $p=0$, the black $p$-branes (7) reduce to the non-dilatonic
 $d$-dimensional black holes. In this case, these critical 
 exponents become
\begin{eqnarray}
&&\alpha=\psi=\gamma=\sigma= 3/2,\ \ \beta=\delta ^{-1}=-1/2,
\nonumber\\
&& \eta=-1, \ \ \bar{d}=1,
\end{eqnarray}
undependent of the dimensionality of spacetime. These 
critical exponents are exactly the same as those of three 
dimensional Ba\~nados-Teitelboim-Zanelli (BTZ) black holes [13]. 
Recall the fact
   that the BTZ black holes are also exact non-dilatonic black hole 
   solutions in string theory [19], we find that these critical 
   exponents are universal for non-dilatonic black holes, an 
   important feature of critical behavior in the
 non-dilatonic black holes.   For the dilatonic black holes 
with the coupling constant $a$ obeying (3), we find
    that the effective spatial dimension is also $p$ 
    (it is one for $p$=0). For a general $a$, the scaling
     laws  still hold, but these 
critical exponents and effective dimension will depend on the 
coupling constant $a$ and the dimension $d$ of spacetime. In this 
case, the effective spatial dimension is
\begin{equation}
\bar{d}=\frac{(d-2)^2a^2}{2(d-3)^2-(d-2)a^2}.
\end{equation}
This statement  is also valid for the dilatonic black 
$p$-branes [14].

Summarizing the above, we have the following conclusions: (1) 
The extremal limit of dilatonic and non-dilatonic black $p$-branes
 is critical point and corresponding critical exponents obey the
  scaling laws.  (2) For the non-dilatonic black holes and black 
  strings, the effective spatial dimension is one. This result 
is also reached in the BTZ black holes and 3-dimensional black 
strings [13]. (3) For the non-dilatonic black $p$-branes (black 
string for $p=1$), the effective dimension is $p$, so does it for
 the dilatonic black holes produced by the double-dimensional
  reduction of the non-dilatonic black $p$-branes. (4) For 
  other dilatonic black holes and  black $p$-branes, 
  the effective spatial dimension  depends on the parameters 
  in theories.  Furthermore, near the extremal limit of the
   non-dilatonic black $p$-branes, from Eqs. (9)-(11), we have
\begin{equation}
S \sim T^{p} ,\ \ M-M_{\rm ext}\sim T^{p+1},
\end{equation}
where $M_{\rm ext}$ is the ADM mass of extremal black $p$-branes.
 Notice that the ADM mass and entropy of black $p$-branes are 
 extensive quantities with respect to the volume of $p$-branes.
  Thus, near the extremal limit, the thermodynamic properties 
  of non-dilatonic black $p$-branes can be described by the 
  blackbody radiation in ($1+p$) dimensions, which also further
   verify that  the effective spatial dimension is $p$. For the 
   dilatonic black holes (4), equation (34) is also valid. Although
    the entropy of dilatonic black holes is not an extensive quantity,
     the entropy can be regarded as the entropy density of the 
     non-dilatonic black $p$-branes. Thus it seems to imply that 
     these dilatonic black hole entropy can also be explained as 
     the way of $p$-branes, although the string coupling becomes 
     very large in this case.

Recall the recent progress in understanding  entropy of 
black holes [5-7], in which the constant dilaton field
 seems to be a necessary condition. Therefore, our conclusions 
 are in complete agreement with the result of these 
 investigations. Further we also give an interpretation
  why the Bekenstein-Hawking entropy may be given a simple world 
  volume interpretation only for the non-dilatonic 
  $p$-branes (including the non-dilatonic black holes). 
  In addition, more recently, Horowitz and Polchinski [20] have
 proposed a 
  correspondence principle, which states that (i) when the 
  size of the black hole horizon drops below the size of a string,
   the typical black hole state becomes a typical state of strings 
   and D-branes with the same charges, and (ii) the mass does not
   change abruptly during the transition. This principle connecting
black holes to weakly coupled strings and D-branes provides a 
statistical  intepretation of entropy of black holes (including 
dilatonic black holes). Therefore, our phase transition in fact 
corresponds to the transition between black hole description and 
string description.

\end{document}